\documentclass[12pt,letterpaper]{article}

\usepackage{amssymb,graphicx,epstopdf}
\usepackage{natbib}
\usepackage{txfonts}
\usepackage{mathptmx}

\newcommand{\beq}{\begin{equation}}
\newcommand{\eeq}{\end{equation}}
\newcommand{\beqa}{\begin{eqnarray}}
\newcommand{\eeqa}{\end{eqnarray}}
\newcommand{\etal}{\emph{et al}}

\title{M Star Astrosphere Size Fluctuations and Habitable Planet Descreening\footnote{To be published in \emph{Astrobiology}}}

\author{\normalsize 
David S.~Smith\footnote{Corresponding author: \texttt{dss@lpl.arizona.edu};
Present affiliation: Lunar and Planetary Laboratory, University of Arizona,
Tucson, AZ 85721}~~and John M.~Scalo\\ \emph{\normalsize Department
of Astronomy, The University of Texas at Austin, Austin, TX 78712}
}

\begin{document}

\maketitle



\begin{abstract} {

Stellar astrospheres---the plasma cocoons carved out of the
interstellar medium by stellar winds---are continually influenced
by their passage through the fluctuating interstellar medium (ISM).
Inside dense interstellar regions, an astrosphere may be compressed
to a size smaller than the liquid-water habitable zone distance.
Habitable planets then enjoy no astrospheric buffering from the
full flux of Galactic cosmic rays and interstellar dust and gas, a
situation we call ``descreening.''  Recent papers (Yeghikyan and
Fahr, Pavlov \etal.) have suggested such global consequences as
severe ozone depletion and glaciation. Using a ram-pressure balance
model that includes gravitational focusing of the interstellar flow,
we compute the size of the astrosphere in the apex direction as a
function of parent star mass.  We derive a dependence on the
parent-star mass $M$ due to gravitational focusing for densities
larger than about $100~(M/M_\odot)^{-2}$ cm$^{-3}$.  We calculate
the interstellar densities required to descreen planets in the
habitable zone of solar- and subsolar-mass stars and find a critical
descreening density of roughly $600~(M/M_\odot)^{-2}$ cm$^{-3}$ for
the Sun's velocity relative to the local ISM. Finally, we estimate
from ISM observations the frequency of descreening encounters as
1--10 Gyr$^{-1}$ for solar-type stars and 10$^2$--10$^9$ times
smaller for M stars.  Given this disparity, we conclude that M star
habitable-zone planets are virtually never exposed to the severe
effects discussed by Yeghikyan and Fahr and Pavlov \etal.

}

\end{abstract}


\section{Introduction}

The heliosphere is the magnetic bubble carved out of the interstellar medium by the expanding solar wind.
A recurring idea is that during the Sun's lifetime encounters
with dense interstellar environments have compressed the heliosphere,
enhancing cosmic ray fluxes and hydrogen accretion, affecting
atmospheric chemistry (including ozone), and perhaps altering surface
mutation rates and climate \citep[e.g.][]{fahr68, mccrea75, talbot+76,
begelman+rees76, talbot+newman77, mckay+thomas78, yabushita+allen89,
bzowski+96, zank+frisch99, scherer+02, florinski+03, yeghikyan+fahr04a,
yeghikyan+fahr04b, pavlov+05a, pavlov+05b, frisch+slavin06, mueller+06, zank+06, mueller+08}.
There may even be empirical evidence for such an encounter in lunar
soil samples \citep{wimmerschweingruber+bochsler00,
wimmerschweingruber+bochsler01}.  A similar variation should appear
for other stars, given that ``astrospheres'' around other stars
have now been directly observed through their interaction with the
interstellar medium \citep{wood+02,wood+05}.

Interest in the potential habitability of planets orbiting stars
of mass smaller than the Sun's---cool, red, faint stars of spectral
class M with masses between 0.08 and 0.6 $M_\odot$---has increased
in recent years.  For simplicity, we refer to all subsolar-mass
stars as M stars, since K main-sequence stars cover such a small
range in mass.  The liquid-water habitable zone (HZ) is defined
essentially by a certain range in received flux at a planet's surface
\citep{kasting+93}, and M star planets have small luminosities, so
the habitable-zone distance $r_\mathrm{HZ}$ can be extremely small,
from 0.3 AU to 0.02 AU as mass varies from 0.6 to 0.08 $M_\odot$.
Habitable planets around M stars may be common for many reasons,
including the large number of M stars compared to any other class
of star, the detection of extrasolar planets orbiting M stars, their
presumed ability to form terrestrial-mass planets if they have
sufficiently massive disks, their observed disk frequency, and even
evidence for debris disk substructure in one case. See \citet{scalo+07}
and \citet{tarter+07} for detailed reviews of these concepts.

The character of the astrospheric environment expected for
habitable-zone planets orbiting stars of different masses has not
been examined.  The present work focuses on the frequency of
reductions in the size of a star's astrosphere to within the HZ or
smaller induced by passage through the fluctuating conditions in
our Galaxy's interstellar medium (ISM).  Such astrosphere size
changes are inevitable given the range of conditions in the ISM and
may result in significant changes in the Galactic cosmic ray (GCR) flux, as scattering
of cosmic rays in the astrosphere is the main filter (besides any
possible planetary magnetosphere) shielding planets from GCRs.  The
effects of GCRs on planetary atmospheres are significant and varied
\citep[e.g.][]{griessmeier+05, grenfell+06}.  In addition,
changes in astrosphere size should lead to large variations in the
planetary accretion rate of interstellar hydrogen or dust, a
possibility that has been discussed for several decades \citep{fahr68,
mccrea75, bzowski+96, yabushita+allen89, florinski+03, yeghikyan+fahr04b,
pavlov+05b}; the climatic changes associated with such accretion
could be varied and severe (e.g. the noctilucent cloud scenario of
\citealt{mckay+thomas78, mckay85}, or ozone depletion as in
\citealt{yeghikyan+fahr04b, pavlov+05b}).

In the present paper we show that the size of astrospheres around
M stars will be controlled by the star's gravity during descreening
episodes and estimate the frequency of these episodes for M stars
as compared to solar-type stars.  Section \ref{sec:models} will
describe the analytic models for astrospheres that we will use,
Section \ref{sec:results} gives the resulting mass and age dependences,
and Section \ref{sec:descreen_freq} will use these models to estimate
the frequency of descreening encounters for solar- and subsolar-mass
stars.

\section{Methods}  \label{sec:models}

\subsection{Astrosphere Model} \label{sec:models_rampress}

\citet{parker63} showed that the location of the termination shock
of the solar wind given by the shock-jump conditions was roughly
where the ram pressure of the outflowing solar wind balanced the
pressure of the counterstreaming ISM: \beq P_\mathrm{ISM}  =
\frac{\rho_0 \varv_0^2 r_0^2}{r_s^2}, \eeq where $r_s$ is the
termination shock distance, $r_0$ is a reference distance, $\rho_0$
is the solar wind mass density at $r_0$, $\varv_0$ is the solar wind
velocity at $r_0$, and $P_\mathrm{ISM}$ is the total pressure of
the ISM, including ram, thermal, and magnetic pressures.  When the
ISM is streaming supersonically and superalfvenically past the
stellar system (such as for the present day solar system), its ram
pressure dominates $P_\mathrm{ISM}$, such that $P_\mathrm{ISM}
\simeq \rho_\mathrm{ISM} V^2$, where $\rho_\mathrm{ISM}$ is the
total ISM density and $V$ is the relative velocity between the star
and the local ISM flow.  Taking the size of the astrosphere $r_a$
to be the distance from the star to the termination shock, we have
\beq \frac{r_a}{r_0} = \left(\frac{n_0}{N}\right)^{1/2}
\frac{\varv_0}{V},\eeq where $n$ and $N$ will henceforth refer to
the total ion number density of the solar wind and ISM, respectively.

To estimate the stellar wind pressure we assume that the wind within
astrospheres of stars of different spectral types (or equivalently
masses) are analogues of the solar wind.  Little information on the
wind is available for stars besides the Sun other than the variation
of mass-loss rate with age and type for a few nearby stars
\citep{wood+02,wood+05}.  The wind speed $\varv_0$ in our model is
taken to be 400 km s$^{-1}$ and assumed to be independent of distance
from the star, as is roughly the case for the solar wind in the
ecliptic plane \citep{whang+03} at radii beyond a few times the
critical point ($\sim$ 0.1 AU).  Recent work suggests that the
steady-state wind may in fact be slower and even subsonic and
subalfvenic near to the star \citep{erkaev+05,preusse+05}, in which
case the interstellar ram pressure would be balanced by wind magnetic
and thermal pressure near M star HZs.  \citet{khodachenko+06},
however, shows that the steadily higher coronal mass ejection rate
on M stars may significantly affect the wind speed in the HZ.  To
examine the possibility of a slow wind near the HZs of the
lowest mass stars, we include a model in our results in which the
wind speed is taken to be 50 km s$^{-1}$ instead of 400 km s$^{-1}$.

For simplicity, we lump the wind density and velocity dependence
on stellar mass and age into a scaling function: \beq\xi(M,t)=
\left(\frac{M}{M_\odot}\right)^\alpha
\left(\frac{t}{4.5~\mathrm{Gyr}}\right)^\beta,\eeq where $\alpha$ and $\beta$
are allowed to vary.  Thus the ram pressure balance equation for
the astrosphere size can be written: \beq \frac{r_a}{r_0} =
\left[\xi(M,t)\frac{n_0}{N}\right]^{1/2} \frac{\varv_0}{V}.\eeq The
indices $\alpha$ and $\beta$ are extremely uncertain as present, but studies
of stellar mass loss by \citet{wood+02,wood+05} suggest that younger
and lower-mass stars may have stronger winds, implying $\alpha\le 0$ and
$\beta \le 0$.  

Observations of \citet{wood+05} suggest $\beta= -2.33$, but, using the
same relation between mass-loss rate and X-ray flux as \citet{wood+05},
the recent surveys of X-ray fluxes of dwarf M stars by
\citet{penz+micela08} suggest $\beta=-1.8$. The mass dependence of the
mass-loss rate is more complicated, however, and observations of
some M stars show a \emph{lower} mass-loss rate than the Sun
\citep{wood+05}. If this is accurate, then it may be possible for
M stars to have a lower mass-loss rate than the Sun, in which case,
they would be easier to descreen. We include this possibility in
our results in an example in which $\alpha=1$.

Though crude, a pure ram pressure balance approximation is surprisingly
accurate at predicting the solar wind termination shock distance
and generally yields heliosphere sizes in the apex direction within
10--20\% of detailed multifluid, multidimensional computational
models \citep[e.g.][]{zank+frisch99, fahr+00, wood+02, florinski+04,
zank+06, mueller+08}.  The obvious weakness of this analytic model
is that, being one-dimensional, it cannot reproduce the full
three-dimensional structure of the global heliosphere. It can,
however, predict the minimum ISM densities above which planets in the
HZ can be descreened for at least part of their orbit.  In one
sense, this model may even overestimate the astrosphere size and
hence the descreening frequency because it neglects charge exchange
with interstellar neutrals.

We have up to now ignored the thermal and magnetic pressure
contributions in the ISM.  Thermal pressure would be important (for
the solar velocity with respect to the local standard of rest) for
ISM densities below about 0.1 cm$^{-3}$, where temperatures are
high, but we are interested in the high densities that could lead
to descreening, and these regions are typically cold \citep[$T \sim
10$--20 K, ][]{ferriere01} compared to stellar velocities.    For thermal
pressure to dominate to ram pressure, the following inequality must
be satisfied: \beq T \gtrsim 3000\, \left(\frac{V}{10\ \mathrm{km\
s}^{-1}}\right)^{2}\ \mathrm{K},\eeq where $V$ is the star-ISM
relative velocity.  For the Sun, $V = 26$ km s$^{-1}$, so it would
need to enter a region with $T\gtrsim$ 20,000 K for the thermal
pressure to compare to the ram pressure in determining the heliosphere
size, and these regions usually have densities well below the regimes
we are considering here. For instance, the warm ISM has a density
below $\sim 0.5$ cm$^{-3}$ and $T \gtrsim 6000$ K \citep{ferriere01}.

Magnetic pressure is likely to be more important, although without
a detailed MHD simulation of the entire ISM structure spanning
orders of magnitude in scale, we can make only a rough estimate of
the effect.  Zeeman observations \citep{crutcher99,bourke+01} can
be expressed in the form of an upper limit to a $B$-density relation:
\beq B(N) = \left\{ \begin{array}{r@{\quad:\quad}l} 3\ \mu\mathrm{G}
& N  < 10\ \mathrm{cm}^{-3}\\ 3\; \left(\frac{N}{10\
\mathrm{cm}^{-3}}\right)^{1/2}\ \mu\mathrm{G} & N  \ge 10\
\mathrm{cm}^{-3} \end{array}\right. . \eeq At densities below 10
cm$^{-3}$ this simply gives a constant pressure $B^2/8\pi$ equal
to $4 \times 10^{-13}$ dyne cm$^{-2}$.  Since the ram pressure is
$8 \times 10^{-12}\ N V_{20}^2$ dyne cm$^{-2}$, the magnetic field
isn't competitive with ram pressure until densities below 0.05
cm$^{-3}$ (e.g. in the Local Interstellar Cloud; \citealt{florinski+04}
have suggested that the magnetic field is anomalously large here).
At higher densities, the square-root scaling gives a pressure that
is proportional to the density, similar to the ram pressure.  In
this case the coefficient of the magnetic pressure acts to effectively
increase the ISM-star velocity in the ram pressure, but the effect
is small, even after multiplying by a small factor \citep{crutcher99}
to statistically account for the fact that only the line-of-sight
component of $B$ is detected by the Zeeman effect.  This is because
typical star-ISM velocities exceed the root-mean-square turbulent
velocity in the ISM, which is in rough equipartition with the
magnetic field.  Given these considerations, we neglect the magnetic
field for simplicity, with the understanding that we somewhat
overestimate the size of the heliosphere at low densities, and that
in more realistic ISM models the magnetic field will undergo large
fluctuations, often uncorrelated with density, that may lead to
episodes of magnetically controlled astrosphere sizes.

\subsection{Gravitational Focusing of the Interstellar Flow}  \label{sec:focus}

As planetary systems traverse dense clouds, gravitational focusing
will determine inflowing particle trajectories.  This was first
pointed out by \citet{fahr68} for the case of collisionless orbits
of protons in the inner solar system and then later first applied
in the present context by \citet{begelman+rees76}.  Previous work
has either neglected gravitational focusing or included its effect
on the star-ISM velocity but only in the fluid limit
\citep[e.g.][]{talbot+76, talbot+newman77}.

Here we assume the ISM flow is collisionless. The mean free path
for Coulomb collisions is $\sim 10^{16}~N^{-1}$ cm, where $N$ is
the local ISM number density, and this is larger than the astrosphere
size in most cases of interest.  This assumption is marginal at low
densities, but at the high densities required for descreening, where
gravitational focusing dominates the accretion, we show below that
the astrosphere size decreases with ISM density faster than does
the collisional mean free path, so the collisionless assumption is
valid.

Neglecting collisions simply means that we neglect fluid
effects, such as gas pressure, and assume the ISM particles follow
ballistic trajectories until they ``collide'' with the wind-ISM
interface, exerting a pressure.  The physics of this interaction
is beyond the scope of this discussion, but ram-pressure balance
approximates the standoff distance between the wind and ISM quite
well, and this is our primary concern.

A complete solution to stellar accretion in the collisionless limit
was derived by \citet{danby+camm57}.  The infall density can be
expressed as an integral that must be evaluated numerically in the
general case, but for our assumption of small gas thermal velocity
dispersion, Danby and Camm find that the density in the apex direction
is \beq\label{eq:dcneqn}
   N = N_0 \frac{(q^2+2p^2)}{2 p (q^2+p^2)^{1/2}},
\eeq where $N_0$  is the unperturbed ISM number density far from
the star, $p \equiv V/\sigma$,  $q^2 \equiv 2 G M/\sigma^2 r$, and
$\sigma$ is the thermal velocity dispersion of the gas.  The ram
pressure on the astrosphere in the apex direction will then be this
density multiplied by \beq\label{eq:vmod}
   V'^2 = V^2 + \frac{2GM}{r},
\eeq where the second term on the right accounts for gravitational
acceleration.  Balancing the outward wind ram pressure with the
inward ISM ram pressure as before yields a quartic polynomial for
the astrosphere size $r_a$ (see \S\ref{sec:resnumer}).  Most of the
mass dependence of our result comes from the importance of the
gravitational focusing terms involving $q^2$ in Eq. \ref{eq:dcneqn}.

\section{Results}\label{sec:results}

\subsection{Descreening in the High-density Limit} \label{sec:reslimit}

The minimum cloud density $N_f$ for focusing to dominate can be
obtained from the condition $b > r_a$, where $b \equiv 2GM/V^2$ is the
distance at which the star's gravity becomes important, and $r_a$ is
the pressure-balance astrosphere size neglecting focusing.  This gives
\beq N_f = 140\ \xi(M,t)\ V_{20}^2\ \left(\frac{M}{M_\odot}\right)^{-2}
\mathrm{cm}^{-3}, \eeq where $V_{20}$ is $V$ in units of 20 km s$^{-1}$.
Thus in dense molecular clouds \citep[$N \sim 10^2$--10$^3$ or larger;
e.g.][]{larson81, scalo85} the accretion flow will be mostly determined by
gravitational focusing except possibly for the very lowest-mass stars.
For reference, the Sun's present speed relative to the local ISM flow
is about 26 km s$^{-1}$ \citep{witte04}.

Assuming that gravitational focusing dominates, the density
at the apex of the astrosphere is \beq N \simeq \frac{N_0}{2
V}\left(\frac{2GM}{r}\right)^{1/2},\eeq where $N_0$ is the ISM density
unperturbed by the star's gravity.  The incoming velocity (much
larger than the star-gas velocity $V$ in this limit) is $V' \simeq
(2GM/r)^{1/2}$.  Expressed in terms of the gravitational radius $b$,
we get \beq N \simeq \frac{N_0}{2} \left(\frac{b}{r}\right)^{1/2}\eeq
and \beq V' \simeq V \left(\frac{b}{r}\right)^{1/2},\eeq where
$b/r \gg 1$ in the gravitational focusing limit.  Equating $N V'^2$
to the stellar wind ram pressure then allows a simple solution for
the astrosphere size of a planetary system when the flow outside the
astrosphere is dominated by gravitational focusing.  The result is \beq
\frac{r_a}{r_0} \simeq 4 \left[\frac{N_0}{\xi(M,t) n_0}\right]^{-2}
\left(\frac{V}{\varv_0}\right)^{-4} \left(\frac{r_0}{b}\right)^3,\eeq or,
substituting for $b$, \beq \frac{r_a}{r_0} \simeq \frac{V^2 \varv_0^4\,
r_0^3}{2\, G^3 M^3} \left[\frac{N_0}{\xi(M,t)\, n_0}\right]^{-2}.
\eeq Notice that in this gravitational focusing limit the astrosphere size
\emph{increases} with increasing stellar velocity $V$.  This unexpected
behavior occurs because the slower the star is moving, the stronger is
the effect of focusing, so that the ram pressure in this case scales like
$V^{-4}$.  And since the astrosphere size varies as $N_0^{-2}$ in this
limit, while the mean free path of the incoming gas to particle-particle
collisions still scales linearly with $N_0^{-1}$, the astrosphere
size is decreasing faster (due to focusing) than the mean free path
is, and the validity of the collisionless approximation, as measured
by the ratio of mean free path to heliosphere size, \emph{increases}
with increasing density.  Finally, the inverse mass dependence of the
astrosphere size means that the collisionless limit becomes rapidly more
accurate for lower mass stars.

By the definition of the HZ (neglecting albedo variations),
$r_\mathrm{HZ} = (L/L_\odot)^{1/2}$ AU where $L_\odot$ is the bolometric
luminosity of the Sun.  Although no single power law can accurately
represent the mass-luminosity relation for low-mass stars (see below), an
adequate relation for our purposes is $L/L_\odot = (M/M_\odot)^\alpha$,
with $\alpha \sim$ 2--3 for masses between 0.1 and 1 $M_\odot$, based
on masses from suitable binary systems \citep{henry+99, delfosse+00},
Hipparcos distances, and empirical bolometric corrections. The
habitable-zone distance is then $r_\mathrm{HZ} = (M/M_\odot)^{\alpha/2}$
AU.  Taking $r_0 = 1$ AU and the average number density of the solar
wind at 1 AU, $n_0$, to be 7 cm$^{-3}$ and evaluating numerical factors:
\beq \frac{r_a}{r_\mathrm{HZ}} \simeq 3.6\times10^5\; \frac{ V_{20}^2\;
[\xi(M,t)]^2}{N_0^2} \left(\frac{M}{M_\odot}\right)^{-3 - \alpha/2}.  \eeq
The minimum ISM density for this descreening condition to hold is \beq
N_d \gtrsim 600\;V_{20}\; \xi(M,t) \left(\frac{M}{M_\odot}\right)^{-3/2 -
\alpha/4} \mathrm{cm}^{-3},\label{eq:ncritapprox}\eeq where the numerical
coefficient is approximate because the star-gas relative velocity will
vary from star to star by factors of a few.   For $\alpha = 2.5$, the
exponent of $M$ is $-2.1$, so a star with $M = 0.5\,M_\odot$ (early
M star) will require an ISM density 4 times larger than the Earth for
complete descreening to occur, while a $0.1\,M_\odot$ star will require a
density about 120 times larger.  Our charge exchange model would yield an
astrosphere a factor of $(\varv_0/V)^{1/2}$ smaller for a given ISM density,
so it would yield a descreening density that is $(\varv_0/V)^{1/4}$ times
smaller than what the ram pressure approach above would yield.  For any
reasonable wind speed and ISM-star relative velocity, this would be only
a factor of 2--3 at most.

This result is easily understood as a combination of several effects.
First, the smaller stellar mass produces less gravitational focusing,
reducing the exterior ram pressure and increasing the astrosphere
size compared to larger masses.  Second, the habitable-zone distance
decreases with decreasing stellar mass, forcing the astrosphere to shrink
to smaller distances for descreening of habitable planets to occur.
Third, the mass- and age-dependent wind scaling factor
$\xi(M,t)$ at a given age probably will increase with decreasing mass,
leading to a stronger wind and larger astrosphere.

So far we have assumed the limiting case in which the density field
is completely dominated by gravitational focusing in order to get
an analytic result.  In the following section, we solve the full
quartic polynomial equation for the density and give the exact (in
the limit of a cold ISM) solution for $r_a/r_\mathrm{HZ}$ and the critical
density for descreening.

\subsection{Descreening in the General Case}
\label{sec:resnumer}

\begin{figure}
  \centering
  \includegraphics[width=0.6\textwidth,angle=270]{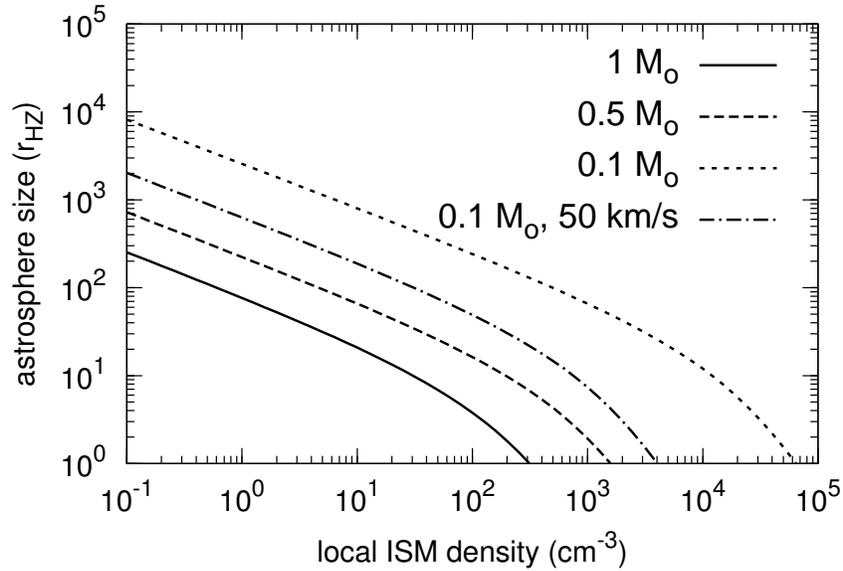} 

  \caption{Astrosphere size, in units of liquid-water habitable-zone
  distance (defined here as the distance receiving the same bolometric
  stellar flux as Earth), as a function of ambient interstellar gas
  density for stars of three different masses.  The regime in which
  the inflowing interstellar ram pressure is dominated by gravitational
  focusing is at densities higher than the ``knee'' in each curve.
  The star-ISM relative velocity was taken to be 10 km s$^{-1}$.
  The stellar wind speed was taken to be 400 km$^{-1}$ in the first
  three cases and 50 km s$^{-1}$ in the last case, representing the
  possibility of subsonic, subalfvenic winds in the HZ of M stars
  \citep{erkaev+05,preusse+05}.  A slower wind speed seems to have
  a significant effect on the astrosphere size, but even in this
  extreme case the astrosphere size is larger for the M star than
  for a typical solar-mass star for all ISM densities.}

  \label{fig:denVApex}

\end{figure}

Following the notation of the previous section, the full polynomial
equation for the astrosphere size can be derived in the collisionless
limit by equating the wind ram pressure with the ISM ram pressure,
using the Danby and Camm (1957) gravitational focusing solution
for the density at the astrospheric apex point and the velocity as
modified by gravitational focusing (Eq. \ref{eq:vmod}).  The result is
\beq\label{eq:fulleqn} r_a^4 + 2\, b\, r_a^3 + \frac{5}{4}\, b^2 r_a^2
+ \frac{1}{4}\, b^3 r_a  - \left[\frac{P_w(M,t)}{N V^2}\right]^2 =
0, \eeq where $b$ is the gravitational radius as defined earlier and
$P_w(M,t) \equiv \xi(M,t)\, n_0 \varv_0^2 r_0^2$.  To solve this equation,
first make the substitution $r_a \equiv u-b/2$.  This results
in the depressed quartic equation: \beq u^4 - {b^2\over 4} u^2 -
\left[\frac{P_w(M,t)}{N V^2}\right]^2  = 0.\eeq  Solving for $u^2$
and taking the positive, real root as the astrosphere size: \beq r_a =
\left( {b^2\over 8} + {1\over 8}\left\{b^4 + \left[\frac{8 P_w(M,t)}{N
V^2}\right]^2 \right\}^{1/2}\right)^{1/2} - {b\over 2}.\eeq The quantity
$r_a/r_\mathrm{HZ}$ as a function of local ISM density is shown for
three representative stellar masses (0.1, 0.5, and 1 $M_\odot$) in
Fig. \ref{fig:denVApex}.

Figure \ref{fig:denVApex} shows that the astrosphere is much farther
away in terms of the habitable-zone distance $r_\mathrm{HZ}$ for
systems with lower-mass parent stars.  The transition into the
gravitationally dominated regime is located, for each of the four
model planetary systems, by the knee in the curves as the ISM density
increases.  This knee occurs for lower ISM densities for systems
with more massive parent stars, since the effect of the star's
gravity on the pressure balance grows at a given ISM density as
mass increases.  One case is shown with a slow, 50 km$^{-1}$, wind
velocity.  This case shows that if M stars do  have weaker winds
than solar-type stars, then they will have correspondingly smaller
astrosphere sizes.  But even in the severe case of a wind speed of
50 km $^{-1}$ shown in Fig.~\ref{fig:denVApex}, the astrosphere is
still larger relative to the HZ distance than for the solar-type
stars.

Figure \ref{fig:denVApex} also shows clearly that an M star must
encounter a very high density interstellar region in order to
compress its astrosphere significantly, and therefore full exposures
to the Galactic cosmic ray and accretion flux will be rare compared
to a solar-type star.  This occurs because denser interstellar
regions occupy a smaller volume fraction of the Galaxy than
lower-density regions, as discussed in \S\ref{sec:descreen_freq}.

\begin{figure}
  \centering \includegraphics[width=0.6\textwidth,angle=270]{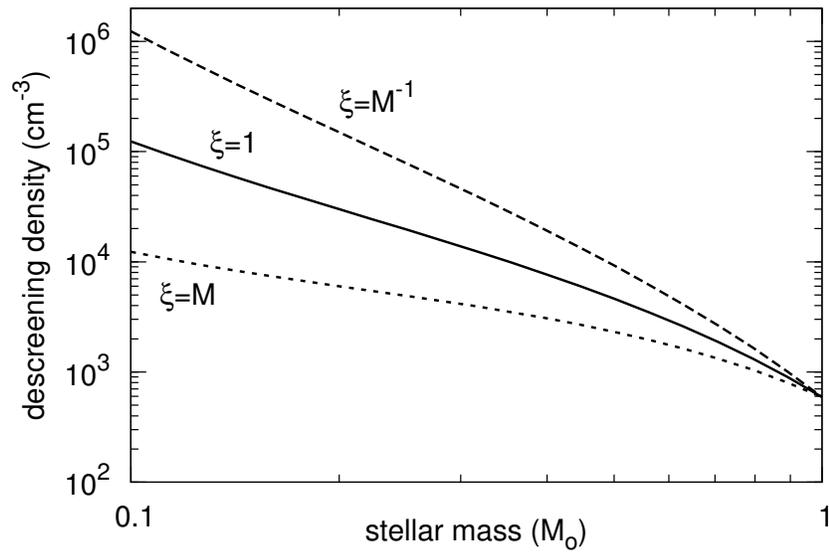}

  \caption{Critical interstellar density for descreening, defined
  as an event during which the astrosphere size is equal to the
  conventional liquid-water habitable-zone distance, as a function
  of parent-star mass in solar units.  Three mass dependences of
  the stellar wind ram pressure are shown.  Comparison with Fig.~1
  shows that descreening densities are at or larger than the density
  above which gravitational focusing dominates. Even in the case
  in which low-mass stars have weaker winds ($\xi \propto M$), the
  descreening density still increases with decreasing mass, due to
  the severe dependence of the HZ distance on stellar mass.}

  \label{fig:nCritVMass}
\end{figure}

A more quantitative calculation of the effect of stellar mass on the
critical descreening density is obtained by solving for the density
at which the astrosphere apex distance shrinks to the habitable-zone
distance $r_\mathrm{HZ}$, and then relating $r_\mathrm{HZ}$ to the stellar
mass.  From Eq. \ref{eq:fulleqn} the critical descreening density is
\beq\label{eq:ncrit} N_d = \frac{P_w(M,t)}{V^2} \left( r_\mathrm{HZ}^4
+ 2\, b\, r_\mathrm{HZ}^3 + \frac{5}{4}\, b^2 r_\mathrm{HZ}^2 +
\frac{1}{4}\, b^3 r_\mathrm{HZ}\right)^{-1/2}.  \eeq The habitable-zone
distance depends on the luminosity of the parent star, $r_\mathrm{HZ}
\sim L^{1/2}$ in Earth-Sun units (again ignoring variations in albedo).
Because the primary (approximately constant) quantity that governs the
evolution of the star is its mass, it is traditional to parameterize
$L$ in terms of stellar mass $M$.  We used the data for stellar masses
and luminosities for stars less massive than 1 $M_\odot$ given in
\citet{hillenbrand+white04} and fit the following polynomial to it:
\beq \log L = 4.10 \log^3 M + 8.16 \log^2 M + 7.11 \log M + 0.065,
\eeq where $L$ and $M$ are in solar units. The slope of this log-log
relation varies from about $2$ at the smallest masses to about $3$
at 0.5--1 $M_\odot$.  This allows us to calculate the critical
descreening density (Eq. \ref{eq:ncrit}) in terms of the parent star mass.
The result is plotted for three different mass dependences of $\xi(M,t)$
in Fig. \ref{fig:nCritVMass} and for an assumed star-ISM velocity of 26
km s$^{-1}$.

A planetary system orbiting a star of mass 1 $M_\odot$ will be
descreened if it encounters a region of density $\gtrsim 600~V_{20}$
cm$^{-3}$.  On the other hand a system orbiting a star of mass 0.2
$M_\odot$ must pass within a region of density $\gtrsim 10^4~V_{20}$
cm$^{-3}$ to be descreened, even if we assume no dependence of wind
strength on mass or age.   If we assume that the star-ISM velocity
parameter, $V_{20}$, scales with the kinematic velocity
dispersion, then $V_{20}$ could be 2--3 times larger for both M
stars \citep{fuchs+09} and older solar-type stars \citep{nordstrom+04}
than for the Sun.  Our results then imply that these stellar systems
would require proportionally larger ISM densities to descreen the
habitable zone.  In this respect, our conclusions are somewhat conservative.

Section \ref{sec:descreen_freq} will argue that such dense regions
will be rarely encountered.

\subsection{Frequency of Descreening as a Function of Stellar Mass}
\label{sec:descreen_freq}

For any reasonable characterization of the ISM, the frequency of
encounters of a planetary system with a cloud of a given density decreases
with increasing density.  Unfortunately this statement is difficult
to quantify because interstellar densities are normally very poorly
determined.  The densities are almost always derived from the observed column
density of some tracer (e.g. dust, or a molecule like CO), a
conversion factor from tracer to to total column density, and then
a conversion to density by dividing the column density by a
characteristic size of the region of interest.  This latter step
is especially uncertain because the apparent size is determined by
the sensitivity limit of the observations (column density normally
decreases outward on average) and by the adopted distance to the
region.  However we can use the approach of \citet{talbot+newman77}
as applied to various catalogues of regions identified as discrete
``clouds'' in different surveys in order to see that, even given these
uncertainties and selection effects, the frequency of occurrence
of dense ISM regions must decrease rapidly with increasing density.

\citet{talbot+newman77} used the \citet{lynds62} catalogue of about 1800 ``dark
clouds,''\,\footnote{Regions on photographic plates that appear
darker than the surrounding area because of reduction in the number of
visible stars by dust extinction.} for which Lynds provided ``opacity
classes'' and angular areas.  Grouping these clouds into bins of a given
column density range using the opacities, and using the mean size $R(N)$
for each group, gives the number of clouds with density in each density
range.  Plotted as a histogram, this gives the number of clouds per unit
volume of space per unit density range, denoted $n_\mathrm{cl}(N)$,
where $N$ is the internal particle density.  If $l(N)$ is the average
number of clouds encountered of density $(N,N+dN)$ per unit line of
sight distance, then $n_\mathrm{cl}(N) = [\pi R(N)^2\; l(N)]^{-1}$.
Since the frequency of encounters is proportional to $l(N)^{-1}$, this
gives a frequency $\nu \propto R(N)^2\; n_\mathrm{cl}(N)$.

We reexamined the cloud property table given by \citet{talbot+newman77},  but we use only the data for the three highest
opacity classes.\footnote{The points for ``Standard cloud,'' actually
derived from reddening statistics, ``CO'' and ``HI'' are ill-defined
averages for entire surveys, and the lower opacity class extinction
clouds suffer from selection effects as explained below and as can
be seen in Fig. 1 of Talbot and Newman.}  Using the derived average
densities in each class, we find the distribution
$n_\mathrm{cl}(N) \propto N^{-1.8}$ and $R(N) \propto N^{-0.8}$,
giving a cumulative frequency of encounters $\nu (>N) \propto N^{-2.4}$.  This
agrees with the exponent of $-2.3$ obtained using directly the $l(>N)$
values in their table, or the scaling of their quantity $N_e$ which
is the number of expected encounters by the Sun over 4.6 Gyr.

This result is very uncertain because the Lynds cloud sample, besides its
small size,  suffers from a number of selection effects:  

1. Clouds with angular sizes less than about 2--5 arcmin cannot be
detected on Palomar Schmidt plates because the number of background
stars is too small. This corresponds to a minimum detectable cloud size
that increases with distance.  Since smaller clouds tend to be denser,
the density probability distribution is probably too flat because of
this effect.

2. Column density detection is somewhat restrictive, because regions
with extinction less than about one magnitude cannot be detected as a
decrease in background star density, while at extinctions above about
5--6 magnitudes clouds look black independent of their extinction.
The latter effect causes the catalogue to lump clouds of densities above
the saturation threshold into the bin of highest column density, again
making the apparent probability distribution of cloud densities appear
too flat.

3. The increasing number of foreground stars makes it progressively
more difficult to detect clouds farther than several hundred parsecs.
This effect produces an underestimate of the extinction to distant large,
and usually lower-density, clouds.

4. Nearby clouds likely obscure more distant clouds, so that some small
clouds located in front of larger clouds are not detected, and so on
\citep{scalo+lazarian96}.

5. Distances to individual clouds detected in extinction are very
uncertain. A way around this would be to examine only clouds in nearby
complexes, like Taurus, which are all at about the same distance,
but then the number of clouds would be small for statistical purposes
\citep[see][]{scalo85}. \citet{talbot+newman77} assumed they were all
at the same distance.

For these reasons we have also derived the \emph{relative} frequency
of encounters with regions above a given density using the FCRAO Outer
Galaxy Survey of carbon monoxide emission, which gives data for over
10,000 clouds \citep{heyer+01}.  This survey is subject to selection
effects analogous to 1 (angular resolution limit) and 2 (CO can only
be excited above a density of about 100 cm$^{-3}$ but the [$^{12}$CO]
line saturates at large column densities; also there is a sensitivity
threshold on column density) above.  The CO sample does not suffer
from uncertainties 3--5 above because the clouds are detected using a
spectral line, and the velocity information not only separates clouds
along the same line of sight but allows distance estimates based on
a kinematic model (rotation curve) for the Galaxy.  We used the cloud
properties from this survey (available electronically) to construct a
density probability distribution $n_\mathrm{cl}(N)$, but chose to include
only clouds with distances less than 5 kpc, since a plot of the number of
clouds vs. distance indicated that incompleteness becomes serious beyond
this point.  We note that the clouds were identified by decomposition
of larger structures, so the catalogue does not contain larger, and
on average less dense, regions within which the catalogued clouds were
contained (Heyer, private communication).  This should not affect our
results much because these larger regions have average densities that
are mostly below the critical descreening density for any but the
most massive parent stars considered here.  Following Heyer \etal.,
we take the calculated CO luminosity $L_\mathrm{CO}$ to be proportional
to the mass of the cloud and then divide by the size cubed in order to
estimate the average internal density.  Following this procedure, we
find that $n_\mathrm{cl}(N)$ declines rapidly with density.  If a power
law is fit to the histogram, excluding the highest and lowest bins,
the cumulative probability distribution function at the high-density
end ($2\times10^3$ to $2\times10^4$ cm$^{-3}$) has a power-law slope of
about $-3$, similar to the result found for the Lynds clouds.  We also
find the $R(N) \propto N^{-1}$ for this sample; this result is in part
a reflection of the same selection effect as for the Lynds cloud sample,
since the surveys tend to be restricted in column density, yielding such
a correlation.  Nevertheless we adopt it as an average correlation.
This gives a density dependence of the cumulative encounter rate of
$N^{-4}$, compared to $N^{-2.4}$ from Talbot and Newman.  We also
performed the same calculation but assuming all the clouds are at the
same distance, to mimic the selection effect due to the unknown distances
in Talbot and Newman's calculation, and find that $n_\mathrm{cl}(>N)$
becomes basically flat, but the distortion should be less severe for
Talbot and Newman because the range in distances is smaller.

We can also estimate the \emph{absolute} frequency of descreening events
for solar-mass stars from the \citet{heyer+01} CO cloud catalogue
and compare to Talbot and Newman's result for extinction clouds.
For the 7900 CO clouds within 5 kpc, we estimate a volume surveyed of
0.86 kpc$^3$, assuming the clouds are distributed in a uniform disk
with a half-thickness of 50 pc, which is close to most estimates for
the scale height of the CO distribution perpendicular to the plane of
the Galaxy. This yields the total cloud space density $n_\mathrm{cl}$.
Most of these clouds are at average internal densities above 300 cm$^{-3}$
(the number declines rapidly below this density, suggesting selection
effects), which is roughly the estimated descreening density for the
solar system (Fig. \ref{fig:nCritVMass}).  The frequency of encounters is
approximately $\nu = \pi \langle R^2\rangle\, n_\mathrm{cl}\, \bar{v} $,
where $\bar{v}$ is the mean star-cloud velocity (see Talbot and Newman,
1977).\footnote{Our frequency estimates assume a random distribution of
clouds, but aren't much changed by the fact that these clouds, as well
as those in Lynds (1962) sample, are actually clustered into larger
structures.  The \emph{mean} encounter frequency is roughly the volume
filling fraction of the clouds divided by their radius, and neither of
these quantities is affected by clustering as long as the clouds don't
overlap.  The time series of encounters will be radically different in
the clustered case, with long periods without descreening punctuated
by clusters of frequent descreenings, but the \emph{average} rate of
encounters is similar if taken over a sufficiently long time interval
($\sim 1$ Gyr here).  For recent history, however, this clustering
effect would be important.  We discuss the details of the time history
during intrusion into a large complex using hydrodynamic simulations in
a separate publication.} We find a mean square cloud radius $\langle
R^2\rangle = 1.2$ pc$^2$. Adopting a mean star-cloud velocity of 20
km s$^{-1}$ gives a descreening frequency of 0.7 Gyr$^{-1}$. This is
about six times smaller than obtained from including only the Lynds
class 3--6 clouds from \citet[][Table 1]{talbot+newman77}.  Since we
found that placing the Heyer \etal. clouds all at the same distance
significantly flattens the $n_\mathrm{cl}$ distribution, the assumption of
a single distance could be part of the discrepancy.  A major difference
arises from the fact that the mean square sizes in the Lynds sample are
much larger than found for the Heyer \etal. sample.  Such differences
are perhaps expected because of the different selection criteria and
selection effects (see \citealt{heyer+01} for discussion in the CO
sample), and the different distances to which the surveys extend, but
this expectation gives us no basis to decide which is the more realistic
estimate. The difference should probably be considered a reflection of
the inherent uncertainty involved in estimating the descreening frequency.
Both results are lower limits because of the resolution effect discussed
earlier, by which smaller, denser clouds are missed at larger distances,
but we do not know the size of the corresponding correction factor.

Given these considerations, we think that the descreening frequency for
the Sun cannot be estimated to better than an order of magnitude, but
probably lies between 1 and 10 Gyr$^{-1}$.  Since the critical descreening
density scales approximately as $M^{-2}$ (Eq.~\ref{eq:ncritapprox}),
and $n_\mathrm{cl}(>N)$ decreases rapidly with $N$, we see that the
descreening frequency declines extremely rapidly with decreasing stellar
mass.  Using the dark cloud $n_\mathrm{cl}(>N)$ power-law slopes found
above, we get a mass dependence of $M^5$, while the CO cloud slopes
yield $M^8$.

It should be pointed out that these results only apply to relatively dense
clouds that can be seen in extinction or CO, and we do not know how the
frequency scales with density for smaller densities.  In addition, our
estimated rates for dense regions are probably underestimates, since
both the Lynds and \citet{heyer+01} cloud catalogues select against
dense clouds because of resolution effects, as discussed above.

Since we showed earlier that an M star planet must encounter a region
of density $\sim 1\times10^3$ cm$^{-3}$ (for M0, $\sim 0.5 M_\odot$)
to $3\times10^4$ cm$^{-3}$ (for M9, $\sim 0.1 M_\odot$) for descreening,
while a solar-type star requires only $\sim 600$ cm$^{-3}$, the frequency
estimates given above imply that descreening will occur 10$^2$ to 10$^9$
times less frequently for M star planets (i.e., never).  Even for somewhat
higher mass stars, the frequency of descreening is apparently such a
strongly decreasing function of density that, given the crude estimate of
1--10 descreening encounters per Gyr for the Earth, even stars of spectral
types late G and early K should be relatively immune from descreening.

\section{Summary and Conclusions}

We have used a ram-pressure balance model for calculating the size
of astrospheres in the apex direction around solar- and subsolar-mass
stars to compute the minimum interstellar density required to
``descreen'' a planet in the habitable zone. We showed that
gravitational focusing of incoming particle trajectories causes an
enhancement of the ISM ram pressure and a consequent dependence of
the astrosphere size on the parent-star mass when the local ISM
density rises above $\sim 100~M^{-2}$ cm$^{-3}$, where $M$ is the
stellar mass in solar units.  We have included the effects of
gravitational focusing in calculating the ISM densities required
to descreen planets in the habitable zone of solar-like and lower-mass
stars and find a critical descreening density of roughly $600~M^{-2}$
cm$^{-3}$.  Finally, we have estimated the frequency of descreening
encounters as 1--10 Gyr$^{-1}$ for solar-type stars and 10$^2$--10$^9$
times less frequent for M stars, so habitable planets around M stars
will most likely never be exposed to the unfiltered ISM flow.

We caution that these results are sensitive to the parameters of
the stellar winds of low-mass stars.  Studies of mass-loss rates
for M stars have not yet well constrained the dependence of the
mass-loss rate on mass or age. Preliminary results
\citep[e.g.][]{wood+02,wood+05} suggest that the wind strength may
increase with decreasing age and mass, but examples to the contrary
do exist.  While we neglect the age dependence of the wind strength
for this discussion, we do find that even a mass dependence in which
the mass-loss rate is linearly dependent on the stellar mass does
not change our results (see Fig.~2).  Deeper surveys of stellar
mass-loss rates as a function of mass and age would be beneficial
to understanding the astrospheric environment 
around low-mass stars, which are potentially the most common sites
for habitable planets in the Galaxy \citep{scalo+07}.

\section*{Acknowledgements}

This work was partially supported by the NASA Exobiology Program,
Grant NNG04GK43G, and the NASA Astrobiology Institute, Virtual Planetary
Laboratory Lead Team.  D.~Smith was supported by the NSF Graduate Student
Research Fellowship and Harrington Doctoral Fellowship Programs.


\begin{thebibliography}{200}

\parindent=0in
\newcommand{\apj}{\emph{Astrophys. J.}}
\newcommand{\aj}{\emph{Astr. J.}}
\newcommand{\apjl}{\emph{Astrophys. J. Lett.}}
\newcommand{\apjs}{\emph{Astrophys. J. Supp.}}
\newcommand{\icar}{\emph{Icarus}}
\newcommand{\sci}{\emph{Science}}
\newcommand{\nat}{\emph{Nature}}
\newcommand{\newa}{\emph{New Astr.}}
\newcommand{\mnras}{\emph{Mon. Not. Roy. Astr. Soc.}}
\newcommand{\araa}{\emph{Ann. Rev. Astr. Astrophys.}}
\newcommand{\aanda}{\emph{Astr. Astrophys.}}
\newcommand{\jgr}{\emph{J. Geophys. Res.}}
\newcommand{\pasp}{\emph{Pub. Astr. Soc. Pacific}}
\newcommand{\oleb}{\emph{Origins Life Evol. Biosphere}}
\newcommand{\pss}{\emph{Plan. Sp. Sci.}}


\bibitem[Abramowitz and Stegun(1972)]{abramowitz+stegun72} Abramowitz, M.,
Stegun, I.A., Eds. (1972) \emph{Handbook of Mathematical Functions}, Dover, NY.

\bibitem[Begelman and Rees(1976)]{begelman+rees76} Begelman, M.C. and
Rees, M.J. (1976) Can cosmic clouds cause climatic catastrophes? \nat,
261, 298--299.

\bibitem[Bourke \etal.(2001)]{bourke+01} Bourke, T.L., Myers, P.C.,
Robinson, G., and Hyland, A.R. (2001) New OH Zeeman measurements of
magnetic field strengths in molecular clouds.  \apj, 554, 916--932.

\bibitem[Bzowski \etal.(1996)]{bzowski+96} Bzowski, M., Fahr, H.J., and
Rucinski, D. (1996) Interplanetary neutral particle fluxes influencing the
Earth's atmosphere and the terrestrial environment. \icar, 124, 209--219.

\bibitem[Crutcher(1999)]{crutcher99} Crutcher, R.M. (1999) Magnetic fields
in molecular clouds: observations confront theory. \apj, 520, 706--713.

\bibitem[Danby and Camm(1957)]{danby+camm57} Danby, J.M.A. and Camm,
G.L. (1957) Statistical dynamics and accretion. \mnras, 117, 50--71.

\bibitem[Delfosse \etal.(2000)]{delfosse+00} Delfosse, X., Forveille, T.,
Segransan, D., Beuzit, J.-L., Udry, S., Perrier, C., and Mayor, M. (2000)
Accurate masses of very low mass stars. IV. Improved mass-luminosity
relations. \emph{Astron. Astrophys.}, 364, 217--224.

\bibitem[Erkaev \etal.(2005)]{erkaev+05} Erkaev, N.V., Penz, T.,
Lammer, H., Lichtnegger, H.I.M., Wurz, P., Biernat, H.K., Griessmeier,
J.-M., and Weiss, W.W. (2005) Plasma and magnetic field parameters
in the vicinity of short periodic giant exoplanets. \apjs, 157,
396--401.

\bibitem[Fahr(1968)]{fahr68} Fahr, H.J. (1968) On the influence of neutral
interstellar matter on the upper atmosphere. \emph{Astrophys. Sp. Sci.},
2, 474--495.

\bibitem[Fahr \etal.(2000)]{fahr+00} Fahr, H.J., Kausch, T., and
Schere, H. (2000) A 5-fluid hydrodynamic approach to model the solar
system-interstellar medium interaction.  \aanda, 357, 268--282.

\bibitem[Ferri\`ere(2001)]{ferriere01} Ferri\`ere, K.M. (2001) The
interstellar environment of our galaxy. \emph{Rev.~Mod.~Phys.}, 73, 1031--1066.

\bibitem[Fite \etal.(1962)]{fite+62} Fite, W., Smith, A., and Stebbings, R.
(1962) Charge transfer in collisions involving symmetric and asymmetric
resonance. \emph{Roy. Soc. London Proc. Ser. A}, 268, 527--536.

\bibitem[Florinski \etal.(2004)]{florinski+04} Florinski, V., Pogorelov,
N.V., Zank, G.P., Wood, B.E. and Cox, D.P. (2004) On the possibility
of a strong magnetic field in the local interstellar medium. \apj,
604, 700--706.

\bibitem[Florinski \etal.(2003)]{florinski+03} Florinski, V., Zank,
G.P., and Axford, W.I. (2003)  The Solar System in a dense interstellar
cloud: Implications for cosmic-ray fluxes at Earth and $^{10}$Be records.
\emph{Geophys. Res. Lett.}, 30, 2206--2210.

\bibitem[Frisch and Slavin(2006)]{frisch+slavin06} Frisch, P.~C., 
Slavin, J.~D.\ 2006.\ The Sun's journey through the local interstellar 
medium: the paleoLISM and paleoheliosphere.\ \emph{Astrophysics and Space 
Sciences Transactions}, 2, 53--61.

\bibitem[Fuchs et al.(2009)]{fuchs+09} Fuchs, B., and 13 others. (2009) The
kinematics of late-type stars in the solar cylinder studied with SDSS data.
\emph{Astron.~J.}, 137, 4149--4159.

\bibitem[Grenfell \etal.(2006)]{grenfell+06} Grenfell, J.L.,
Griessmeier, J.-M., Patzer, B., Rauer, H., Segura, A., Stadelmann,
A., Stracke, B.,  Titz, R., von Paris, P. (2006) Biomarker Response
to Galactic Cosmic Ray Induced NO$_x$ and the Methane Greenhouse Effect
in the Atmosphere of an Earthlike Planet Orbiting an M Dwarf Star.
\emph{Astrobiology}, 7, 208--220.


\bibitem[Griessmeier \etal.(2005)]{griessmeier+05} Griessmeier,
J.-M., Stadelmann, A., Motschmann, U., Belivsheva, N.K., Lammer,
H., Biernat, H.,K.  (2005) Cosmic ray impact on Extrasolar Earth-like
planets in close-in habitable zones. \emph{Astrobiology}, 591,
1--12.

\bibitem[Henry \etal.(1999)]{henry+99} Henry, T.J., Franz, O.G.,
Wasserman, L.H., Benedict, G.F., Shelus, P.J., Ianna, P.A., Kirkpatrick,
J.D., and McCarthy, D.W. (1999) The optical mass-luminosity relation at
the end of the main sequence (0.08--0.20 $M_\odot$). \apj, 512,
864--873.

\bibitem[Heyer \etal.(2001)]{heyer+01} Heyer, M.H., Carpenter, J.M.,
and Snell, R.L. (2001) The equilibrium state of molecular regions in
the outer Galaxy. \apj, 551, 852--866.

\bibitem[Hillenbrand and White(2004)]{hillenbrand+white04}
Hillenbrand, L.A. and White, R.J. (2004) An assessment of dynamical mass
constraints on pre-main-sequence evolutionary tracks. \apj, 604, 741--757.

\bibitem[Kasting \etal.(1993)]{kasting+93} Kasting, J.F., Whitmire,
D.P., and Reynolds, R.T. (1993) Habitable zones around main sequence
stars. \icar, 101, 108--128.

\bibitem[Khodachenko \etal.(2006)]{khodachenko+06} Khodachenko,
M.L., and 10 others. (2006) Coronal Mass Ejection (CME) Activity
of Low Mass M Stars as an Important Factor for the Habitability of
Terrestrial Exoplanets. I. CME Impact on Expected Magnetospheres of
Earth-like Exoplanets in Close-in Habitable Zones.  \emph{Astrobiology},
7, 167--184.

\bibitem[Larson(1981)]{larson81} Larson, R.B. (1981) Turbulence and
star formation in molecular clouds. \mnras, 194,
809--826.


\bibitem[Lynds(1962)]{lynds62} Lynds, B.T. (1962) Catalogue of dark
nebulae.  \apjs, 7, 1--60.

\bibitem[McCrea(1975)]{mccrea75} McCrea, W.H. (1975) Ice ages and the
Galaxy. \nat, 255, 607--609.

\bibitem[McKay(1985)]{mckay85} McKay, C.P. (1985) Noctilucent cloud
formation and the effects of water vapor variability on temperatures in
the middle atmosphere. \emph{Planet. Space Sci.}, 33. 761--771

\bibitem[McKay and Thomas(1978)]{mckay+thomas78} McKay, C.P. and Thomas,
G.E. (1978) Consequences of a past encounter of the Earth with an
interstellar cloud.  \emph{Geophys. Res. Lett.}, 5. 215--218.

\bibitem[M{\"u}ller et al.(2006)]{mueller+06} M{\"u}ller,
H.-R., Frisch, P.~C., Florinski, V., Zank, G.~P.\ 2006.\ Heliospheric
Response to Different Possible Interstellar Environments.\ \apj,
647, 1491--1505.


\bibitem[M{\"u}ller et al.(2008)]{mueller+08} M{\"u}ller, H.-R.,
Frisch, P.~C., Fields, B.~D., and Zank, G.~P.\ (2008) The Heliosphere
in Time. Sp.~Sci.~Rev., 163.

\bibitem[Nordstr\"om et al.(2004)]{nordstrom+04} Nordstr\"om, B., Mayor, M.,
Andersen, J., Holmberg, J., Pont, F., J{\o}rgensen, B.R., Olsen, E.H., Udry, S.,
and Mowlavi, N. (2004) Ages, metallicities, and kinematic properties of
$\sim$ 14,000 F and G dwarfs. \aanda, 418, 989--1019.


\bibitem[Parker(1963)]{parker63} Parker, E.N. (1963)  \emph{Interplanetary
dynamical processes}, Wiley, NY, pp. 113--128.

\bibitem[Pavlov \etal.(2005a)]{pavlov+05a} Pavlov, A.A., Pavlov,
A.K., Mills, M.J., Ostryakov, V.M., Vasilyev, G.I., and Toon, O.B.
(2005a) Catastrophic ozone loss during passage of the Solar system
through an interstellar cloud. \emph{Geophys. Res. Lett.}, 32,
L01815--L01818.

\bibitem[Pavlov \etal.(2005b)]{pavlov+05b} Pavlov, A.A., Toon, O.B.,
Pavlov, A.K., Bally, J., and Pollard, D.  (2005b) Passing through a giant
molecular cloud: ``Snowball'' glaciations produced by interstellar dust.
\emph{Geophys. Res. Lett.}, 32,  L03705--L03708.

\bibitem[Penz and Micela(2008)]{penz+micela08} Penz, T., and Micela,
G.  (2008) X-ray induced mass loss effects on exoplanets orbiting
dM stars.  \aanda, 479, 579--584.

\bibitem[Preusse \etal.(2005)]{preusse+05} Preusse, S., Kopp, A.,
B\"uchner, J., and Motschmann, U. (2005) Stellar wind regimes of
close-in extrasolar planets. \emph{Astron. Astrophys.}, 434,
1191--1200.

\bibitem[Scalo(1985)]{scalo85} Scalo, J.M. (1985) \emph{Fragmentation and
hierarchical structure in the interstellar medium.} In \emph{Protostars
and Planets II}, ed. D.C. Black and M.S. Matthews, Univ. Ariz. Press,
pp. 201--296.

\bibitem[Scalo \etal.(2007)]{scalo+07} Scalo, J.M., and 14 others. (2007) M
Stars as Targets for Terrestrial Exoplanet Searches and Biosignature
Detection. \emph{Astrobiology}, 7, 85--166.

\bibitem[Scalo and Lazarian(1996)]{scalo+lazarian96} Scalo, J.M. and
Lazarian, A. (1996) Occlusion effects and the distribution of interstellar
cloud sizes and masses. \apj, 469, 189--193.

\bibitem[Scherer \etal.(2002)]{scherer+02} Scherer, K.,
Fichtner, H., and Stawicki, O. (2002) Shielded by the wind: the
influence of the interstellar medium on the environment of the
Earth. \emph{J. Atmospher. Solar-Terr. Phys.}, 64, 795--804.

\bibitem[Smith \etal.(2003)]{smith+03} Smith, E.J., Marsden, R.G.,
and 10 others. (2003) The sun and heliosphere at solar maximum.  \sci,
302, 1165--1169.

\bibitem[Stone \etal.(2005)]{stone+05} Stone, E.C., Cummings, A.C.,
McDonald, F.B., Heikkila, B.C., Lal, N., Webber, W.R. (2005) Voyager 1
explores the termination shock region and the heliosheath beyond. \sci,
309, 2017--2020.


\bibitem[Talbot and Newman(1977)]{talbot+newman77} Talbot, R.J. and
Newman, M.J. (1977) Encounters between stars and dense interstellar
clouds. \apjs, 34, 295--308.

\bibitem[Talbot \etal.(1976)]{talbot+76} Talbot, R.J., Butler, D.M.,
and Newman, M.J. (1976) Climatic effects during passage of the Solar
System through interstellar clouds.  \nat, 262, 561--562.

\bibitem[Tarter \etal.(2007)]{tarter+07} Tarter, J.C., and 31 others. (2007)
A Reppraisal of the Habitability of Planets around M Dwarf Stars.
\emph{Astrobiology}, 7, 30--65.

\bibitem[Whang \etal.(2003)]{whang+03} Whang, Y.C., Burlaga, L.F., Wang,
Y.-M., and Sheeley, N.R. (2003) Solar wind speed and temperature outside
10 AU and the termination shock.  \apj, 589, 635--643.

\bibitem[Wimmer-Schweingruber and
Bochsler(2000)]{wimmerschweingruber+bochsler00} Wimmer-Schweingruber,
R.F. and Bochsler, P. (2000) Is there a record of interstellar pick-up
ions in lunar regolith? In \emph{Acceleration and Transport of Energetic
Particles Observed in the Heliosphere: AIP Conf. Proc.},  edited by
R.A. Mewaldt \etal., Melville, NY, pp. 270--273.

\bibitem[Wimmer-Schweingruber and
Bochsler(2001)]{wimmerschweingruber+bochsler01} Wimmer-Schweingruber,
R.F. and Bochsler, P. (2001) A non-solar origin of the ``SEP'' component
in lunar soils.  In \emph{The Outer Heliosphere: The Next Frontiers,
COSPAR Colloquia Ser. 11}, edited by K. Scherer, H. Fichtner, H.J. Fahr,
and E. Marsch, Pergamon Press, Amsterdam, pp. 507--510.

\bibitem[Witte(2004)]{witte04} Witte, M. (2004) Kinetic parameters
of interstellar neutral helium: Review of results obtained during one
solar cycle with the Ulysses/GAS-instrument. \emph{Astron. Astrophys.},
426, 835--844.

\bibitem[Wood \etal.(2002)]{wood+02} Wood, B.E., M\"uller, H.-R., Zank,
G.P., and Linsky, J.L. (2002) Measured mass-loss rates of solar-like
stars as a function of age and activity.  \apj, 574, 412--425.

\bibitem[Wood \etal.(2005)]{wood+05} Wood, B.E., M\"uller, H.-R., Zank,
G.P., Linsky, J.L., and Redfield, S. (2005) New Mass-Loss Measurements from
Astrospheric Ly$\alpha$ Absorption. \apj, 628, L143--L146.

\bibitem[Yabushita and Allen(1989)]{yabushita+allen89} Yabushita, S. and
Allen, A.J. (1989) On the effect of accreted interstellar matter on the
terrestrial environment. \mnras, 238, 1465--1478.

\bibitem[Yeghikyan and Fahr(2004a)]{yeghikyan+fahr04a} Yeghikyan, A. and
Fahr, H. (2004a) Effects induced by the passage of the Sun through dense
molecular clouds. I. Flow outside of the compressed heliosphere. \aanda,
415, 763--770.

\bibitem[Yeghikyan and Fahr(2004b)]{yeghikyan+fahr04b} Yeghikyan,
A. and Fahr, H. (2004b) Terrestrial atmospheric effects induced by
counterstreaming dense interstellar cloud material. \aanda, 425,
1113--1118.

\bibitem[Zank(1999)]{zank99} Zank, G.P. (1999) Interaction of
the solar wind with the local interstellar medium: A theoretical
perspective. \emph{Sp. Sci.  Rev.}, 89, 412--688.

\bibitem[Zank and Frisch(1999)]{zank+frisch99} Zank, G.P. and Frisch,
P.C. (1999) Consequences of a change in the Galactic environment of the
Sun. \apj, 518, 965--973.

\bibitem[Zank et al.(2006)]{zank+06} Zank, G.~P., M{\"u}ller, 
H.-R., Florinski, V., Frisch, P.~C.\ 2006.\ Heliospheric Variation in 
Response to Changing Interstellar Environments.\ Solar Journey: The 
Significance of our Galactic Environment for the Heliosphere and Earth 338, 
23. 

\end{thebibliography}
\end{document}